\documentstyle[12pt]{article}
\begin{document}

\thispagestyle{empty}

\baselineskip=0.6cm

\noindent P.~N.~Lebedev Institute Preprint     \hfill
FIAN/TD/13--93\\ I.~E.~Tamm Theory Department       \hfill
\begin{flushright}{September 1993}\end{flushright}

\begin{center}

\vspace{0.5in}

{\Large\bf ON THE MULTILEVEL GENERALIZATION }

\bigskip

\vspace{0.3in}

{\Large\bf OF THE FIELD -- ANTIFIELD FORMALISM }

\bigskip

\vspace{0.3in}
{\large  I.~A.~Batalin and I.~V.~Tyutin}\\
\medskip  {\it Department of Theoretical Physics} \\ {\it  P.~N.~Lebedev
Physical Institute} \\ {\it Leninsky prospect, 53, 117 924, Moscow,
Russia}$^{\dagger}$\\

\end{center}

\vspace{1.5cm}

\centerline{\bf ABSTRACT}
\begin{quotation}

The multilevel geometrically--covariant generalization of the
field--antifield BV--formalism is suggested. The structure of quantum
generating equations and hypergauge conditions is studied in details. The
multilevel formalism is established to be physically--equivalent to the
standard BV--version.

\end{quotation}

\vfill

\noindent

$^{\dagger}$ E-mail address: tyutin@uspif.if.usp.br

\newpage

\setcounter{page}{2}

\newpage

\section{Introduction}

In previous paper [1] of the present authors a generalization
of the field--antifield BV--formalism [2--4] has been suggested in which
all the field and antifield variables are treated on equal footing to
coordinatize the extended phase space.

The most characteristic feature of the new formalism is that not only the
quantum master equation but also the hypergauge conditions are formulated
without making use of explicit field--antifield splitting. The corresponding
hypergauge functions are not quite arbitrary but satisfy some equations, the
so--called unimodular involution relations. These relations are formulated
in a geometrically--covariant way and thus do not destroy the
field--antifield uniformity.

It has been shown in Ref. [1] that the unimodular involution relations can be
obtained as a result of a further generalization of the formalism. Namely,
one should extend the original phase space by including the antifields
conjugated to the hypergauge Lagrangian multipliers. In the extended phase
space one constructs the second--level hypergauge theory, being the original
hypertheory called the first--level one. In special hypergauge the
second--level theory can be reduced to the first--level one in such a way
that the first--level hypergauge appears to satisfy automatically the
unimodular involution relations required.

In its own turn, the second--level hypertheory can be extended further to
become the third--level one, and so on.

A natural hypothesis appears that there exists a unified hypertheory that
includes the Lagrangian multipliers and their conjugated antifields of all
the levels, together with the corresponding chain of hypergauge conditions.
Regrettably, at the present stage we are able to formulate the hypertheory
of a fixed level only, being the final--level hypergauge condition imposed
by hand.

In the present paper we construct explicitly the fixed--level hypertheory
and study in details the structure of hypergauge conditions, that provides
for a gauge invariance of the formalism.

As is usual, we denoted by $\varepsilon(A)$ the Grassmann parity of a quantity
$A$.

Other notation is clear from the context.

\section{ Outline of the first--level formalism }

Let $\Gamma^A$, $A=1,\ldots,2N$, $\varepsilon(\Gamma^A)\equiv\varepsilon_A$,
be a total set of field--antifield variables coordinatizing the original phase
space.

We define the antisymplectic differential $\Delta$ to be a general
second--order fermionic operator without the derivativeless term,

$$\Delta={1\over2}(-1)^{\varepsilon_A}M^{-1}\partial_AME^{AB}\partial_B,
\eqno{(2.1)}$$
required to satisfy the nilpotency condition, $\Delta^2=0$, so that
$E^{AB}(\Gamma)$ appears to be antisymplectic metric satisfying the Jacobi
identity and thus yielding the antibracket operation:

$$(F,G)\equiv F\overleftarrow{\partial_A}E^{AB}\overrightarrow{\partial_B}G
{}.
\eqno{(2.2)}$$

The first--level functional integral is defined as follows:

$$Z=\int\!\!\exp\{{\imath\over\hbar}[W(\Gamma;\hbar)+G_a(\Gamma)\pi^a]\}d\mu,
\eqno{(2.3)}$$
where

$$d\mu=Md\Gamma d\pi \eqno{(2.4)}$$
is the integration measure, the action $W(\Gamma;\hbar)$ satisfies the
quantum master equation

$$\Delta\exp\{{\imath\over\hbar} W(\Gamma;\hbar)\}=0, \eqno{(2.5)}$$
$\pi^a$, $a=1,\ldots,N$, $\varepsilon(\pi^a)\equiv\varepsilon_a$, are the
Lagrangian multipliers introducing the hypergauge conditions fixed by the
functions $G_a$ that satisfy the so--called unimodular involution relations:

$$(G_a,G_b)=G_cU^c_{ab}, \eqno{(2.6)}$$

$$\Delta G_a-U^b_{ba}(-1)^{\varepsilon_b}=G_bV^b_a, \eqno{(2.7)}$$

$$V^a_a=G_a\tilde{G}{}^a, \eqno{(2.8)}$$
with $U^c_{ab}$, $V^b_a$, $\tilde{G}{}^a$ to be some functions of the
original phase space variables $\Gamma$.

The integrand of eq. (2.3) is invariant under the generalized BRST--type
transformations:

$$\delta\Gamma^A=(\Gamma^A,-W+G_a\pi^a)\mu, \eqno{(2.9)}$$

$$\delta\pi^a=(-U^a_{bc}\pi^c\pi^b(-1)^{\varepsilon_b}+2\imath\hbar V^a_b\pi^b+
2(\imath\hbar)^2\tilde{G}{}^a)\mu,\eqno{(2.10)}$$
where $\mu=\hbox{const}$, $\varepsilon(\mu)=1$.

Choosing the parameter $\mu$ to be function

$$\mu={\imath\over2\hbar}X(\Gamma) \eqno{(2.11)}$$
that satisfies the equations:

$$\imath\hbar\Delta X=G_aK^a,\quad\Delta(G_aK^a)=0,
\eqno{(2.12)}$$
and making the additional variations

$$\delta\Gamma^A={1\over2}(\Gamma^A,X),\quad\delta\pi^a=K^a, \eqno{(2.13)}$$
one generates the following change of the hypergauge functions $G_a$ alone

$$\delta G_a=(G_a,X) \eqno{(2.14)}$$
in the functional integral (2.3).

\section{ The $n$-th--level formalism }

In this Section we construct inductively the $n$-th--level functional
integral for $n=2,3,\ldots$.

Let us define recursively the $n$-th--level set of variables of the
field--antifield phase space:

$$\Gamma^{(n)A_{(n)}}\equiv( \Gamma^{(n-1)A_{(n-1)}}; \pi^{(n-1)a},
\pi^{*(n-1)}_a ) , \eqno{(3.1)}$$
where

$$\Gamma^{(1)A_{(1)}}\equiv\Gamma^A,\quad\pi^{(1)a}\equiv\pi^a. \eqno{(3.2)}$$
with $\pi^{(n-1)a}$ and $\pi^{*(n-1)}_a$ to be $(n-1)$-th--level Lagrangian
multipliers and their conjugated antifields, respectively, so that

$$\varepsilon(\pi^{(n)a})=\varepsilon(\pi^{*(n)}_a)+1=\varepsilon_a
+n-1. \eqno{(3.3)}$$

Further, one constructs the operator $\Delta^{(n)}$ :

$$\Delta^{(n)}=\Delta^{(n-1)}+\Delta^{(n)}_\pi, \eqno{(3.4)}$$

$$\Delta^{(n)}_\pi=(-1)^{(\varepsilon_a+n)}
{\partial_l\over\partial\pi^{(n-1)a}}{\partial_l\over\partial\pi^{*(n-1)}_a},
\eqno{(3.5)}$$

$$\Delta^{(1)}\equiv\Delta. \eqno{(3.6)}$$

Let us assign to the n-th level, $n\ge2$, the corresponding Planck constant
$\hbar^{(n)}$, $\varepsilon(\hbar^{(n)})=0$, in addition to the usual one
$\hbar$, together with the new quantum number called the Planck parity
$\hbox{Pl}^{(n)}$:

$$\hbox{Pl}^{(n)}(\Gamma^{(n-1)})=\hbox{Pl}^{(n)}(\hbar)=0, \eqno{(3.7)}$$

$$\hbox{Pl}^{(n)}(\hbar^{(n)})=\hbox{Pl}^{(n)}(\pi^{(n-1)})=
-\hbox{Pl}^{(n)}(\pi^{*(n-1)})=1. \eqno{(3.8)}$$

The $n$--th--level quantum action $W^{(n)}(\Gamma^{(n)};\hbar;\hbar^{(n)})$ is
defined to satisfy the quantum master equation:

$$\Delta^{(n)}\exp\{{\imath\over\hbar^{(n)}}
W^{(n)}(\Gamma^{(n)};\hbar;\hbar^{(n)})\}=0. \eqno{(3.9)}$$

The action $W^{(n)}$ possesses the quantum numbers:

$$\varepsilon(W^{(n)}(\Gamma^{(n)};\hbar;\hbar^{(n)}))=0,\quad
\hbox{Pl}^{(n)}(W^{(n)}(\Gamma^{(n)};\hbar;\hbar^{(n)}))=1, \eqno{(3.10)}$$
and has the following series expansion in powers of $\hbar^{(n)}$,
$\pi^{(n)}$, $\pi^{*(n)}$ :

$$W^{(n)}(\Gamma^{(n)};\hbar;\hbar^{(n)})=
\Omega^{(n)}(\Gamma^{(n)};\hbar)+\imath\hbar^{(n)}
\Xi^{(n)}(\Gamma^{(n)};\hbar)+
(\imath\hbar^{(n)})^2\tilde{\Omega}{}^{(n)}(\Gamma^{(n)};\hbar)
+\ldots, \eqno{(3.11)}$$

$$\begin{array}{c}
\Omega^{(n)}(\Gamma^{(n)};\hbar)=G_a^{(n-1)}(\Gamma^{(n-1)};\hbar)
\pi^{(n-1)a}+ \\[9pt]
+{1\over2}\pi^{*(n-1)}_cU_{ab}^{(n-1)c}(\Gamma^{(n-1)};\hbar)
\pi^{(n-1)b}\pi^{(n-1)a}(-1)^{(\varepsilon_a+n)}+\ldots,
\end{array} \eqno{(3.12)}$$

$$\Xi^{(n)}(\Gamma^{(n)};\hbar)=-{\imath\over\hbar}W^{(n-2)}
(\Gamma^{(n-2)};\hbar)+\pi^{*(n-1)}_aV_b^{(n-1)a}(\Gamma^{(n-1)};\hbar)
\pi^{(n-1)b}+\ldots, \eqno{(3.13)}$$

$$\tilde{\Omega}^{(n)}(\Gamma^{(n)};\hbar)=\pi^{*(n-1)}_a
\tilde{G}^{(n-1)a}(\Gamma^{(n-1)};\hbar)+\ldots, \eqno{(3.14)}$$
where

$$W^{(n)}(\Gamma^{(n)};\hbar)\equiv
W^{(n)}(\Gamma^{(n)};\hbar;\hbar),\quad n\ge 2, \eqno{(3.15)}$$

$$W^{(1)}(\Gamma^{(1)};\hbar)\equiv W,\quad W^{(0)}\equiv 0,
\quad G^{(1)}_a\equiv G_a. \eqno{(3.16)}$$

The $n$-th--level functional integral is defined as follows:

$$Z^{(n)}=\int\!\!\exp\{{\imath\over\hbar}[W^{(n-1)}(\Gamma^{(n-1)};\hbar)
+W^{(n)}(\Gamma^{(n)};\hbar)+G^{(n)}_a(\Gamma^{(n)})\pi^{(n)a}]\}d\mu^{(n)},
 \eqno{(3.17)}$$

$$d\mu^{(n)}=d\mu^{(n-1)}d\pi^{*(n-1)}d\pi^{(n)},\quad n\ge2, \eqno{(3.18)}$$

$$Z^{(1)}\equiv Z,\quad d\mu^{(1)}\equiv d\mu. \eqno{(3.19)}$$

The $n$--th--level hypergauge functions should satisfy the relations
\footnote {All the antibrackets, $(\, ,\, )$, are always understood to include
the totally--extended set of field--antifield variables} :

$$(G^{(n)}_a,G^{(n)}_b)=G^{(n)}_cU^{(n)c}_{ab}, \eqno{(3.20)}$$

$${\imath\over\hbar}(W^{(n-1)}(\Gamma^{(n-1)};\hbar),G^{(n)}_a)+
\Delta^{(n)}G^{(n)}_a+U^{(n)b}_{ba}(-1)^{(\varepsilon_b+n)}=
G^{(n)}_bV^{(n)b}_a, \eqno{(3.21)}$$

$$V^{(n)a}_a=G^{(n)}_a\tilde{G}{}^{(n)a}, \eqno{(3.22)}$$
with some functions $U^{(n)a}_{bc}$, $V^{(n)a}_b$, $\tilde{G}{}^{(n)a}$.
Besides, some normalization conditions should be imposed on $G^{(n)}_a$, to
be considered in Section 5.

The following remark is relevant here. Being the $n$-th--level theory under
consideration, the functions $G^{(n)}_a$ are subordinated to the relations
(3.20) -- (3.22) by hand, while the preceding functions $G^{(k)}_a$,
$\tilde{G}{}^{(k)a}$, $U^{(k)a}_{bc}$, $V^{(k)a}_b$, $1\le k\le n-1$, are to
be found by solving the equations for $W^{(k)}$, $2\le k\le n-1$. Besides,
all the functions $G^{(k)}_a$, $1\le k\le n-1$, are restricted by
normalization conditions analogous to the ones imposed on the functions
$G^{(n)}_a$.

It will be shown below that the functional integral $Z^{(n)}$ does not
depend on the choice of $G^{(n)}_a$, and coincides, in special hypergauge,
with the $(n-1)$-th--level functional integral $Z^{(n-1)}$, being the
functions $G^{(n-1)}_a$ to fix the hypergauge in this integral.

\section{ Gauge invariance of the $n$-th--level formalism }

In this Section we show the functional integral to be $G^{(n)}$--independent
and equivalent to $Z^{(n-1)}$.

The integrand of (3.17) is invariant under transformations:

$$\delta\Gamma^{(n)A_{(n)}}=(\Gamma^{(n)A_{(n)}},W^{(n-1)}-W^{(n)}+
G^{(n)}_a\pi^{(n)a})\mu^{(n)}, \eqno{(4.1)}$$

$$\delta\pi^{(n)a}=[U^{(n)a}_{bc}\pi^{(n)c}
\pi^{(n)b}(-1)^{(\varepsilon_b+n)}
+2\imath\hbar V^{(n)a}_b\pi^{(n)b}+2(\imath\hbar)^2
\tilde{G}{}^{(n)a}]\mu^{(n)}. \eqno{(4.2)}$$
where $\mu^{(n)}=\hbox{const}$, $\varepsilon(\mu^{(n)})=1$.

Choosing the parameter $\mu^{(n)}$ to be function

$$\mu^{(n)}={\imath\over2\hbar}X^{(n)} \eqno{(4.3)}$$
that satisfy the equation

$$\imath\hbar\Delta^{(n)}X^{(n)}-(W^{(n-1)},X^{(n)})=
G^{(n)}_aK^{(n)a}, \eqno{(4.4)}$$
and making the additional variations

$$\delta\Gamma^{(n)A_{(n)}}={1\over2}(\Gamma^{(n)A_{(n)}},X^{(n)}),
\quad\delta\pi^{(n)a}=K^{(n)a}, \eqno{(4.5)}$$
one generate the following change of the hyperfunctions alone

$$G^{\prime(n)}=G^{(n)}_a+\delta G^{(n)}_a,\quad\delta G^{(n)}_a=
(G^{(n)}_a,X^{(n)}) \eqno{(4.6)}$$
in functional integral (3.17).

The transformations (4.6) retains the form of the unimodular involution
relations \\ (3.20) -- (3.22) by inducing the following transformation of the
structure functions:

$$U^{\prime(n)a}_{bc}=U^{(n)a}_{bc}+(U^{(n)a}_{bc},X^{(n)}), \eqno{(4.7)}$$

$$V^{\prime(n)a}_b=V^{(n)a}_b+(V^{(n)a}_b,X^{(n)})+(-1)^{(\varepsilon_a+n)}
U^{(n)b}_{ac}K^{(n)c}+(-1)^{(\varepsilon_a+n)(\varepsilon_b+n)}
(G^{(n)}_a,K^{(n)b}), \eqno{(4.8)}$$

$$\begin{array}{c}
\tilde{G}{}^{\prime(n)a}=\tilde{G}{}^{(n)a}+(\tilde{G}^{(n)a},X^{(n)})+
V^{(n)a}_bK^{(n)b}+\\[9pt]
+(-1)^{(\varepsilon_a+n)}(W^{(n-1)},K^{(n)a})-
(-1)^{(\varepsilon_a+n)}\Delta^{(n)}K^{(n)a}.
\end{array} \eqno{(4.9)}$$

It will be shown in the next Section that the variation (4,6) induces the most
general actual changes admissible for the hypergauge surface $G^{(n)}_a=0$.
Thus we conclude that $Z^{(n)}$ does not depend on hypergauge fixing.

Let us suppose the functions $G^{(n)a}$ to be solvable with respect to the
antifields $\pi^{*(n)}_a$,

$$\hbox{Sdet}\biggl|{\partial_l G^{(n)}_a\over\partial\pi^{*(n-1)}_b}
\biggr|{\raise-6pt \hbox{$|_{G^{(n)}_a=0}$}}\neq0. \eqno{(4.10)}$$

Choosing the simplest hypergauge

$$G^{(n)}_a=\pi^{*(n)}_a \eqno{(4.11)}$$
of the class (4.10), one reduce $Z^{(n)}$ to the form:

$$Z^{(n)}=\int\!\!\exp\{{\imath\over\hbar}[W^{(n-2)}
+W^{(n-1)}+G^{(n-1)}_a\pi^{(n-1)a}]\}d\mu^{(n-1)},  \eqno{(4.12)}$$

To identify this representation with the $(n-1)$-th--level functional
integral $Z^{(n-1)}$, it is sufficient to show the functions $G^{(n-1)}_a$ to
satisfy the relations of the form (3.20) -- (3.22). Substituting the expansion
(3.11) for $W^{(n)}$ into the quantum master equation (3.9), we find the
following equations for the functions $\Omega^{(n)}$, $\Xi^{(n)}$,
$\tilde{\Omega}{}^{(n)}$, $n\ge2$ :

$$(\Omega^{(n)},\Omega^{(n)})=0, \eqno{(4.13)}$$

$$(\Omega^{(n)},\Xi^{(n)})=\Delta\hspace{-0,05cm}^{(n)}\hspace{0.05cm}
\Omega^{(n)}, \eqno{(4.14)}$$

$$(\Omega^{(n)},\tilde{\Omega}{}^{(n)})=
\Delta\hspace{-0,05cm}^{(n)}\hspace{0.05cm}\Xi^{(n)}-
{1\over2}(\Xi^{(n)},\Xi^{(n)}). \eqno{(4.15)}$$
To the lowest orders in $\pi^{(n-1)}$, $\pi^{*(n-1)}$ these equations give:

$$(G^{(n-1)}_a,G^{(n-1)}_b)=G^{(n-1)}_cU^{(n-1)c}_{ab}, \eqno{(4.16)}$$

$${\imath\over\hbar}(W^{(n-2)},G^{(n-1)}_a)+
\Delta^{(n-1)}G^{(n-1)}_a+U^{(n-1)b}_{ba}(-1)^{(\varepsilon_b+n-1)}=
G^{(n-1)}_bV^{(n-1)b}_a, \eqno{(4.17)}$$

$$V^{(n-1)a}_a=G^{(n-1)}_a\tilde{G}{}^{(n-1)a}, \eqno{(4.18)}$$

Thus the quantum master equation yields automatically the relations required
to use $G^{(n-1)}_a$, as well as the lower $G^{(k)}_a$, to be the
hypergauge fixing functions.

\section{ Structure of hypergauge conditions }

In this section we consider in details the structure of hypergauge functions
that follows from relations (4.16) -- (4.18), including the transformation
properties, normalization conditions, and equivalence between the simplest
hypergauge $G^{(n)}_a=\pi^{*(n-1)}_a$ and arbitrary ones.

In what follows we mean the $n$-th--level hypertheory, $n=2,\ldots$. The
label $n$ will be omitted. For instance, $W$, $\pi$, $\pi^*$, $\Gamma$,
$\Delta$ denoted $W^{(n-1)}$, $\pi^{(n-1)}$, $\pi^{*(n-1)}$,
$\Gamma^{(n-1)}$, $\Delta^{(n)}$, respectively. the case $n=1$ has been
considered in Ref. [1].

As the integrand of $Z$ contains the hypergauges $G_a$ inside the
$\delta$--function only, we are only interested in their properties on the
hypergauge surface

$$G_a=0. \eqno{(5.1)}$$

Let these equations possess the solution

$$\varphi_a\equiv\pi^*_a-f_a(\Gamma,\pi)=0. \eqno{(5.2)}$$

Let us expand $G_a$ in $\varphi$--power series:

$$G_a(\Gamma,\pi,\pi^*)=\varphi_b\Lambda^b_a(\Gamma,\pi)+
\varphi_c\varphi_b\Lambda^{bc}_a(\Gamma,\pi)+\ldots. \eqno{(5.3)}$$
The only essential for the integration of $Z$ are $\Lambda^a_b$ and
$\varphi_a$. Substituting the expansion (5.3) into the relations (3.20),
(3.21), we obtain the following equations for $\varphi_a$ and $\Lambda^b_a$ :

$$(\varphi_a,\varphi_b)=0, \eqno{(5.4)}$$

$$(D,\varphi_a)=-Q\varphi_a, \eqno{(5.4)}$$

$$Q\equiv{\imath\over\hbar}W+\Delta,\quad Q^2=0,\qquad
D\equiv\ln\hbox{Sdet}\Lambda^a_b. \eqno{(5.6)}$$

One can solve these equations explicitly:

$$f_a(\Gamma,\pi)=-E(\hbox{ad}\Psi){\partial_l\over\partial\pi^a}\Psi,
\eqno{(5.7)}$$

$$D=-E(\hbox{ad}\Psi)Q\Psi+\exp(\hbox{ad}\Psi)d_0(\Gamma), \eqno{(5.8)}$$
where $\Psi(\Gamma,\pi)$ and $d_0(\Gamma)$ are arbitrary functions,

$$E(x)={\exp(x)-1\over x}, \eqno{(5.9)}$$
and operator ad$\Psi$ acts according to the rule:

$$\hbox{ad}\Psi A\equiv(\Psi,A). \eqno{(5.10)}$$

Let us expand $X$ in $\varphi$--power series:

$$X=X_0(\Gamma,\pi)+\varphi_aM^a(\Gamma,\pi)+\ldots. \eqno{(5.11)}$$
Substituting this expansion into the equation (4.4), we see that this
equation imposes no restrictions on $X_0$, so that $X_0$ appears to be an
arbitrary function.

Let us consider the transformation properties of $f_a$ and $D$ under the
hypergauge variations.

By making use of (4.6), (5.4) -- (5.6) and the relation:

$$G_a+\delta G_a=(\varphi_b+\delta\varphi_b)(\Lambda^b_a+\delta\Lambda^b_a)+
(\varphi_c+\delta\varphi_c)(\varphi_b+\delta\varphi_b)
(\Lambda^{bc}_a+\delta\Lambda^{bc}_a), \eqno{(5.12)}$$
we find:

$$\delta\varphi_a=(\varphi_a,X_0), \eqno{(5.13)}$$

$$\delta D=(D,X_0)+QX_0. \eqno{(5.14)}$$
These transformation properties are described by the following variations of
arbitrary functions:

$$\delta\Psi=-E^{-1}(\hbox{ad}\Psi)X_0, \eqno{(5.15)}$$

$$\delta d_0=0. \eqno{(5.16)}$$

It follows from (5.15), (5.16) that the correspondence between $\delta\Psi$
and $X_0$ is one--to--one. That means that arbitrary $\Psi$ can be
transformed to become zero by using a finite hypergauge transformation. Gauge
invariance of $d_0$ means that this function can be normalized. Our choice of
the normalization condition is:

$$d_0\equiv0, \eqno{(5.17)}$$
so that

$$D=-E(\hbox{ad}\Psi)Q\Psi. \eqno{(5.18)}$$
Thus an arbitrary set of hypergauge functions $G_a$ can be transformed to
become of the form:

$$G_a=\pi^*_b\Lambda^b_a(\Gamma,\pi)+O((\pi^*)^2), \eqno{(5.19)}$$

$$\hbox{Sdet}\Lambda^b_a=1. \eqno{(5.20)}$$
As not the matrix $\Lambda^b_a$ itself but only its superdeterminant enters
the expression for $Z$, the set (5.19) is equivalent to the one

$$G_a=\pi^*_a. \eqno{(5.21)}$$

\end{document}